\documentstyle[12pt,a4,psfig]{article}
\newcommand{\be}{\begin{equation}}
\newcommand{\ee}{\end{equation}}
\newcommand{\ba}{\begin{eqnarray}}
\newcommand{\ea}{\end{eqnarray}}
\newcommand{\NN}{\nonumber}

\newcommand{\rcite}[1]{{\cite{#1}}}
\newcommand{\rref}[1]{{(\ref{#1})}}
\newcommand{\tref}[1]{{\ref{#1}}}
\newcommand{\fref}[1]{{ Fig.\ref{#1}}}
\newcommand{\rlabel}[1]{{\label{#1}}}
\newcommand{\rbibitem}[1]{\bibitem{#1}}

\newcommand{\tr}{\mbox{tr}}

\def\pr#1, #2, #3 {Phys.~Rev.~\underbar{D#1}, #2~(19#3)}
\def\prc#1, #2, #3 {Phys.~Rep.~\underbar{C#1}, #2~(19#3)}
\def\prl#1, #2, #3 {Phys.~Rev.~Lett.~\underbar{#1}, #2~(19#3)}
\def\pl#1, #2, #3 {Phys.~Lett.~\underbar{B#1}, #2~(19#3)}
\def\mpla#1, #2, #3 {Mod.~Phys.~Lett.~\underbar{A#1}, #2~(19#3)}
\def\ijmp#1, #2, #3 {Int.~Jour.~Mod.~Phys..~\underbar{A#1}, #2~(19#3)}
\def\np#1, #2, #3 {Nucl.~Phys.~\underbar{B#1}, #2~(19#3)}
\def\Bnp#1, #2, #3 {~\underbar{B#1}, #2~(19#3)}

\begin{document}
\begin{titlepage}
\begin{flushright}
hep-ph/9607462
\end{flushright}
\vspace{1.0cm}
\begin{center}
{\Large\bf Electromagnetic Contributions to
 Vector Meson Masses and Mixings} \\[1cm]
J. Bijnens and  P. Gosdzinsky\\[0.5cm]
NORDITA, Blegdamsvej 17\\
DK-2100 Copenhagen \O , Denmark\\
\end{center}
\vfill
\begin{abstract}
We use the $1/N_c$ method to estimate electromagnetic contributions
to vector meson masses and mixings. We identify several new sources
of $\rho-\omega$ mixing coming from short-distance effects.
We comment on the extraction of quark masses from the vector meson masses.
We also present a simple group theoretical discussion of the electromagnetic
mass differences.
\end{abstract}
\vfill
July 1996
\end{titlepage}
\section{Introduction}
The discussion of vector meson masses and mixings has a long history.
The early history can be found in the description of currents by vector meson 
dominance\rcite{sakurai}. The effect of other electromagnetic effects on 
$\rho-\omega$ mixing was estimated to be small\rcite{gourdin}. In this letter
we will estimate these effects on $\rho-\omega$ mixing and on the
individual masses. We will use the expansion in the numbers of colours ($N_c$)
\rcite{tHooft} as an organizing principle throughout.

The leading term is ${\cal O}(N_c)$ and is the mixing of $\rho$, $\omega$ and
$\phi$ with the photon. This is well known and is reviewed shortly
in section \tref{mixing}. At the next-order in $1/N_c$ or ${\cal O}(1)$ there
also appear short distance contributions. Here we perform the long and short
distance matching using the $1/N_c$ method of Buras et al.\rcite{buras}.
This was then extended to include proper identification of the matching scale
in the calculation of the $\pi^+-\pi^0$ electromagnetic mass 
difference\rcite{bbg}. In this method the PCAC relation to two-point functions
was used. This cannot be extended beyond the chiral limit and for
mesons other than the pseudoscalar octet. A method that allows this
extrapolation was then applied to the $K^+-K^0$ electromagnetic mass difference
where in fact a large violation of Dashen's rule was found\rcite{dashen}. 
Variations of this method have also been used for the $K^0-\overline{K^0}$ mass
difference\rcite{bgk} and the kaon $B_K$ parameter\rcite{bp}.
Here we apply this 
method to the masses of vector mesons. The long distance part of ${\cal O}(1)$
in the $1/N_c$ expansion we estimate simply by using cut-off photon loops.
We present an argument based on heavy vector meson chiral perturbation theory
to show that this is the leading contribution\rcite{wise}. We perform this 
calculation both in the relativistic formalism and in the heavy vector meson
formalism. A technical point regarding ``heavy quark'' integrals with an
Euclidean cut-off is discussed in the appendix.


\section{Effective Lagrangians and a phenomenological analysis of the
vector masses}
\rlabel{lagrangians}

First we describe the relativistic formalism we use and then the heavy
vector meson one. We will use a notation which is more appropriate for
the $1/N_c$ limit than the references indicated.  The various
relativistic versions of describing vector mesons are all equivalent
\rcite{ecker} and are all connected by field redefinitions. For the one
connecting the antisymmetric tensor representation with the others see
\rcite{equiv} and the others can be found in \rcite{ecker}.

The vector mesons are collected in a matrix $V_\mu$ with
\be
V_\mu = \left(
\begin{array}{ccc}
\frac{1}{\sqrt{2}}\rho^0_\mu + \frac{1}{\sqrt{2}}\omega_\mu & \rho^+_\mu
 & K^{*+}_\mu \\
\rho^-_\mu & -\frac{1}{\sqrt{2}}\rho^0_\mu+\frac{1}{\sqrt{2}}\omega_\mu &
 K^{*0}_\mu \\
K^{*-}_\mu & \overline{K^{*0}_\mu} & \phi_\mu
\end{array}\right)\,.
\ee
The pseudoscalar mesons are similarly collected in a three by 
three matrix $M$ with
\be
M = \left(
\begin{array}{ccc}
\frac{1}{\sqrt{2}}\pi^0 + \frac{1}{\sqrt{6}}\eta _8 & \pi^+
 & K^{+} \\
\pi^- & -\frac{1}{\sqrt{2}}\pi^0 +\frac{1}{\sqrt{6}}\eta _8 &
 K^{0} \\
K^{-} & \overline{K^0} & -\frac{2}{\sqrt{6}}\eta _8
\end{array}\right)\,.
\ee
we do not include the $\eta'$ here since that one is heavy due to the
breaking of the relevant axial symmetry by the anomaly.
The relativistic vector meson lagrangian we will use corresponds to
model II in \rcite{ecker} with Lagrangian
\be
{\cal L}_R = -\frac{1}{4}\tr\left(D_\mu V_\nu - D_\nu V_\mu\right)^2
+\frac{1}{2} m_V^2 \tr V_\mu V^\mu +\cdots\,.
\ee
The ellipses denote terms we do not use here.
From the covariant derivative we only need the term coupling to the photon
\be
D_\mu V_\nu = \partial_\mu V_\nu - i e \left[ Q , V_\nu \right ] +\cdots\,,
\ee
with $e$ the unit of charge and $Q$ the quark charge matrix, $Q =
\mbox{diag}( 2/3, -1/3, -1/3)$.

The heavy vector meson chiral lagrangian\rcite{wise} 
(see also \rcite{eichten} and \rcite{georgi}) is given by
\be
{\cal L}_H = -i \tr \left(W^\dagger_\mu v\cdot D W^\mu\right)
+ i g \tr\left(\{W^\dagger_\mu , W_\nu \} u_\alpha\right) v_\beta
\epsilon ^{\mu\nu\alpha\beta}
+ a \tr\left(\chi_+  \{W^\dagger_\mu , W^\mu \} \right)
\rlabel{heavyL}
\ee
where we have only kept the terms leading in $1/N_c$.
The matrix $W_\mu$ is the heavy meson version of $V_\mu$ and $W^\dagger_\mu$
its hermitian conjugate since now $W_\mu$ only contains
the annihilation operators while $W^\dagger_\mu$ contains
the creation operators. $v$ is the velocity of the heavy meson and
$D_\mu W_\nu = \partial_\mu W_\nu + [\Gamma_\mu , W_\nu]$. The matrices
$\Gamma_\mu$, $u_\mu$ and $f^{(+)\mu\nu}$ are defined in \rcite{ecker}.
Here we need only
\be
\Gamma_\mu = -i e Q A_\mu\,,\quad f^{(+)}_{\mu\nu} = 2 e Q F_{\mu\nu}
\quad\mbox{and}\quad u_\mu = -\frac{\sqrt{2}}{F}\partial_\mu M\,.
\ee
The meson mass has disappeared out of \rref{heavyL}. 
Only meson mass differences
appear in the heavy meson Lagrangian. In principle a term
$\mu\, \tr W^\dagger_\mu W^\mu$
exists but is removed by the choice of heavy fields. A singlet octet
mass difference would be produced by a term
$d\, \tr W^\dagger_\mu \tr W^\mu$ and similar terms with more traces
exists for the other terms in \rref{heavyL}. Their effects on the 
electromagnetic mass differences are small but we take them into account in the
photon-vector-meson mixing by using the physical values of the $\rho^0$, 
$\omega$ and $\phi$ masses and mixings with the photon. We also take them
into account in the first phenomenological estimate given below. So
there we add
\be
d\, \tr W^\dagger_\mu \tr W^\mu
+ b\, \left(\tr\left(\chi_+  W^\dagger_\mu\right) \tr\left( W^\mu  \right)
  +\mbox{h.c.}\right)
+ c\, \tr\chi_+ \tr \{W^\dagger_\mu , W^\mu \} \,.
\ee
The $c$ term shifts all vector meson masses by a similar amount and will
not be discussed further. The $b$ like term is small (see bellow), 
since it is higher order in a combined chiral and $1/N_c$ expansion,
but will be kept. A term
like $\tr W^\dagger_\mu \tr(v\cdot D W^\mu)$ treated as a $1/N_c$ correction
can be removed using the equations of motion from \rref{heavyL}.

In the limit where the quark masses vanish the leading two orders in $1/N_c$
of the electromagnetic mass difference can be described
by an effective lagrangian
\be
\delta_1\tr\left( Q W^\dagger_\mu\right)\tr\left(Q W^\mu\right)+
\delta_2\tr\left([ Q ,W^\dagger_\mu]\, [Q, W^\mu]\right)+
\delta_3\tr\left( Q^2 \{W^\dagger_\mu ,W^\mu\}\right)\,.
\ee
There are no terms like $\tr\left( Q^2 W^\dagger_\mu\right)\tr W_\mu$ to the 
order we are working.
Notice the difference with the pseudoscalar electromagnetic masses where
only the combination $\tr\left(Q U Q U^\dagger\right)$ appears to leading order,
which is like the $\delta_2$ term. 

The meson masses and mixings obtained from these lagrangians are :
\ba
\Delta\rho^+ \!\!\! &=& \!\!\!-\delta _2+\frac{5}{9} \delta _3+
4 Ba(m_u+m_d)+8 Bc(m_u+m_d+m_s) \NN \\
\Delta\rho^0 \!\!\! &=& \!\!\! \frac{\delta _1}{2}+\frac{5}{9}\delta _3+
4Ba(m_u + m_d) + 8 Bc(m_u+m_d+m_s) \NN \\
\Delta\rho\omega \!\!\! &=& \!\!\! \frac{\delta _1}{6}+
\frac{\delta _3}{3}
+4 Ba(m_u - m_d)+4 Bb(m_u -m_d) \NN \\
\Delta\rho\phi \!\!\! &=& \!\!\! -\frac{ \delta _1}{3 \sqrt{2}}
 + 2\sqrt{2} Bb(m_u - m_d) \NN \\
\Delta\omega \!\!\! &=& \!\!\! 2 d +\frac{\delta _1}{18} 
+\frac{5}{9}
 \delta _3 + 4 Ba(m_u + m_d) +8 Bb(m_u + m_d) \NN \\
& +& \!\!\! 8Bc(m_u + m_d + m_s) \NN \\
\Delta\omega\phi \!\!\! &=& \!\!\! \sqrt{2} d
 -\frac{\delta _1}{9\sqrt{2}} 
+2\sqrt{2}Bb(m_u + m_d + 2m_s) \NN \\
\Delta\phi \!\!\! &=& \!\!\! d + \frac{\delta _1}{9} +
 \frac{2}{9}\delta _3 +8 Bam_s + 8Bbm_s
+8 Bc(m_u+m_d+m_s) \NN  \\
\Delta K^{*+} \!\!\! &=& \!\!\!- \delta _2 + \frac{5}{9}\delta _3 
+ 4 B a(m_s + m_u) + 8 Bc(m_u+m_d+m_s) \NN \\
\Delta K^{*0} \!\!\! &=&  \!\!\! \frac{2}{9} \delta _3 +
 4 B a(m_s + m_d)
+ 8 Bc(m_u+m_d+m_s)
\rlabel{masses}
\ea
The relations
\ba
0 \!\!\! &=& \!\!\!  m_\phi + m_{\rho^+} - m_{K^{*+}}  - m_{K^{*0}} 
-\sqrt{2}\Delta \omega \phi +\frac{ m_\omega - m_{\rho ^0} }{2} \NN \\
0 \!\!\! &=& \!\!\! \Delta \rho \omega - \sqrt{2} \Delta \rho \phi -
m_{ K^{*+}} + m_{ K^{*0}}  -
m_{ \rho ^0} + m_{\rho ^+}
\rlabel{relation}
\ea
can easily be obtained.

We can now try to determine all these parameters from experiment.
Using \rcite{gasrep}
\be
|\Delta \rho\omega |^2 = \left[ (m_\rho-m_\omega)^2 +\frac{1}{4}(\Gamma_\rho
 -\Gamma_\omega)^2\right] \frac{\Gamma(\omega\to\pi\pi)}{\Gamma(\rho\to\pi\pi)}
\ee
and the branching ratio of $\omega\to\pi\pi$ of $2.11\pm 0.30\%$\rcite{PDG}
we obtain
\be
\Delta\rho\omega = -2.5~\mbox{MeV}\,.
\ee
where the sign is determined from the interference pattern of $\rho^0$ and 
$\omega$ in $e^+ e^-\to\pi^+\pi^-$.
Similar estimates can made of the $\omega - \phi$ and the $\rho - \phi$
mixing
using the branching ratios in $3\pi$ and $2\pi$ respectively. Here there
can be very large corrections from Kaon rescattering into
these final states
so these numbers should not be regarded as very informative. We obtain
\be
| \Delta\rho\phi | = 0.33\pm0.1~\mbox{MeV}
\qquad\mbox{and}\qquad | \Delta\omega\phi | = 
8.1~\mbox{MeV}\,.
\rlabel{expmixings}
\ee
We can also estimate the $\omega \phi$ mixing
from the decays $ \omega ,\phi \rightarrow \pi ^0 \gamma$.
We find
\be
| \Delta\omega\phi | = 14.1~ \mbox{MeV}.
\rlabel{expmixingbis}
\ee
Using these numbers and the measured masses we see that the first
relation is well satisfied but the second one is not. This already indicates
that there will be higher order corrections.

A small technical note. We calculate here the inverse of the vector meson 
propagator and determine its zero. In the relativistic case we take
only the $g_{\mu\nu}$ part of this inverse propagator for simplicity:
\be
i g_{\mu\nu}\left(-p^2 + M^2 + \Delta(p^2)\right)  =0 .
\ee
To the order we are working we now have
\be
\Delta M^2 = \Delta(p^2 = M^2)\,.
\ee
Similarly in the heavy meson case,
\be
i\left( -v\cdot p + \Delta M(p^2,v.p)\right) = 0.
\ee
Here the mass shift is given by
\be
\Delta M = \Delta M(0,0) = \frac{\Delta M^2}{ 2M }\,.
\ee

\section{Long distance : Mixing with the photon}
\rlabel{mixing}

We can add to the Lagrangian \rref{heavyL} a term describing the mixing with
the photon. This term has the form
\be
\frac{i f_V \sqrt{m _V} }{2} 
\tr\left([ e^{-i{m_V}\,v\cdot x} W^\dagger_\nu - 
e^{i{m_V}\,v\cdot x}W_\nu] v_\mu
 f^{(+)\mu\nu}\right)
\rlabel{photonmixing}
\ee
The factor $\sqrt{m_V}$ is present because of the normalization
of the $W_\mu$
field. This can be obtained from the relativistic Lagrangian
\be
-\frac{f_V}{2\sqrt{2}} \tr\left( (D_\mu V_\nu - D_\nu V_\mu ) 
f^{(+)\mu\nu}\right)\,,
\ee
by inserting the relation between $W_\mu$ and $V_\mu$, and keeping only
leading terms in $1/m_V$.
The parameter $f_V$ can be estimated from the decays into $e^+ e^-$ using
\be
\Gamma\left[(\rho^0,\omega,\phi)\to e^+e^-\right] = 
\left(1,\frac{1}{9},\frac{2}{9}\right)\frac{4\pi \alpha^2}{3}
 m_V f_V^2\,.
\rlabel{vectoree}
\ee
This leads to $|f_V|=(0.2,0.18,0.16)$ MeV, showing good agreement between the
three observed decays. The diagram in \fref{diagrams}.e leads
to a mass difference of the form described by the $\delta_1$ term
and is given by
\be
\delta_1 =  4 \pi \alpha f_V^2 m_V .
\ee
If we used the relativistic formulation we would have obtained the same
mass differences but with $m_V$ replaced by the relevant vector mass.
This is one possible estimate of ${\cal O}(m_q)$ corrections.

\section{Long distance : Photon loops}
\rlabel{photonloops}
Here we only have to compute the diagrams of \fref{diagrams}.a and
\fref{diagrams}.b.
The calculation is by rotating the integrals in Euclidean space and then
cutting off on the off-shellness of the photon momentum.
In the relativistic case both diagrams contribute and we obtain

\begin{eqnarray}
\delta m_V^2 \!\!\! &=& \!\!\!  -{i \alpha \over 4 \pi^3} \int d^4 k
\left\{ \left( { 1\over k^2 + 2 p\cdot k} \right)
         \left( {1 \over 3} -{k^2 \over 3 m_V^2} +4{m_V^2 \over k^2}
         \right) + {1\over 3 m_V^2} + {1\over k^2} \right\} \nonumber \\
\!\!\! &+& \!\!\! 3{i \alpha \over 4 \pi ^3} \int d^4 k {1\over k^2} 
\rlabel{fullmass}
\end{eqnarray}
for the case of the charged $\rho$ and $K^*$
and zero for the others. The first line of \rref{fullmass} is the contribution
of \fref{diagrams}.a, and the second line is the contributions of the
tadpole, \fref{diagrams}.b.  
We have checked that the gauge dependence of the photon propagator cancels.
We can evaluate all the relevant
integrals and obtain

\ba
\delta ^{LD}  m_V^2 \!\!\! &=& \!\!\! {\alpha m_V^2 \over \pi}
\left\{ {1\over 16} { \Lambda^4 \over m_V^4} +
{1 \over 72 } { \Lambda^6 \over  m_V^6 } +
{11 \over 6} \log  {\Lambda (1+ \sqrt{ 1 + 4 m_V ^2 / \Lambda ^2} )
\over 2 m_V } \right. \nonumber  \\
\!\!\! &-&\!\!\!
\left. { \Lambda ^2 \over 3 m_V^2} \sqrt{ 1 + 4 m_V^2 / \Lambda^2}
\left( -{13 \over 8} + {5 \over 48} {\Lambda^2 \over m_V ^2}
+ {1\over 24} {\Lambda ^4 \over m_V ^4} \right) \right\}
\ea

This is a long distance calculation so we have $\Lambda$ below about 1~GeV.
We can also take the approximation of $\Lambda$ small and obtain
\be
\Delta M^2 = \frac{\alpha}{\pi} \left( 2  m_V \Lambda + O(\Lambda^2) 
\right)
\ee
Both expressions are plotted in \fref{plot} for $\Lambda$ between
500 and 800 MeV.

Alternatively, we could directly have used the heavy meson lagrangian
\rref{heavyL}. Then only the diagram in \fref{diagrams}.c exists
and we obtain
\be
\Delta M = \frac{\alpha \Lambda}{\pi} 
\rlabel{hmetmass}
\ee
The relevant integral is evaluated in the appendix with an euclidean cutoff
and we see that it directly reproduces the small $\Lambda$ result of
the relativistic version. We can also see here that in order to obtain 
contributions that match the short distance of the neutral mesons we have to
go beyond the lowest order in the chiral expansion.
This is similar to what was observed in \rcite{dashen}
where the lowest order electromagnetic loop also gave no mass to the neutral
particles where as the short distance one did.

This contribution can be summarized as
\be
\delta_2^{LD} = - \frac{\alpha}{\pi}\Lambda
\ee

That it is the leading term in the combined chiral and $1/N_c$ expansions
can be seen from the Lagrangian \rref{heavyL}. Loops with pseudoscalar
mesons are higher order in the chiral expansion.

\section{Short distance contribution}
\rlabel{short}
Here we take the short distance effective action as evaluated within this
approach in \rcite{dashen}. There were three terms there. The first one
describes the electromagnetic shifts in the value of the underlying quark
masses and it's effect can be absorbed in them. We therefore do not consider
it any more. The Box and the Penguin-like contributions are given by
\ba
S_{eff}^2 \!\!\! &=& \!\!\!\frac{\alpha \alpha _S}{27 \Lambda ^2}
\left\{4 \bar u u + \bar d d + \bar s s \right\} _{V_{\alpha \beta} }
\left\{ \bar u u + \bar d d + \bar s s\right\} _{V_{\beta \alpha} } \\
S_{eff}^3  \!\!\! &=& \!\!\! \frac{-\alpha \alpha _S}{6 \Lambda ^2}
\left\{ \phantom{\bar d} \hspace{-3mm}
 -4(uu) - (dd) - (ss) + 4(ud) + 4(us) - 2(ds) \right\}
\rlabel{SSD}
\ea
up to terms of order $m_q^2$. Here $\{ \bar q q \} _{V_{\alpha \beta} }
 = \bar q _\alpha \gamma _\mu q _\beta $ and
$(q q') = (\bar q _\alpha \gamma _\mu \gamma _5 q _\beta)
(\bar q' _\beta \gamma _\mu \gamma _5 q' _\alpha)$.
 It now remains to estimate the matrix elements
of these operators between vector meson states.
In leading order in $1/N_c$ we can Fierz the operators in \rref{SSD}.
The matrix elements are then in leading $1/N_c$
\ba
\langle \rho^0_1 | \bar{u}_\alpha(\gamma_5) \gamma_\mu u_\beta
 \bar{u}_\beta (\gamma_5)  \gamma^\mu u_\alpha | \rho^0_2\rangle
\!\!\!  &=& \!\!\!
-(+)2\langle\bar{u}u\rangle \langle \rho^0_1 | \bar{u}u| \rho^0_2\rangle 
+  \langle\rho^0_1|\bar{u}\gamma_\mu u|0\rangle
\langle 0 | \bar{u}\gamma^\mu u|\rho^0_2\rangle\nonumber\\
\!\!\! &=& \!\!\! 
\left(-(+)8 B_0^2 F^2 B_0 a + \frac{1}{2}f_V^2 m_V^3\right)\epsilon_1\cdot
\epsilon_2\,.
\rlabel{example}
\ea
This can then be evaluated from \rref{heavyL} since the
relevant couplings to external currents are there and
leads to the second line in \rref{example}. All other matrix elements
can be obtained in a similar way.
The relevant formulae in the chiral limit for 
$\delta_2^{SD}$ and $\delta_3^{SD}$ are
\ba
\delta_2^{SD} \!\!\! &=& \!\!\!
\frac{3 \alpha \alpha _S}{4 \Lambda ^2} f_V^2 m_V^3  \NN  \\
\delta_3^{SD} \!\!\! &=&  \!\!\!
\frac{\alpha \alpha _S}{12 \Lambda ^2} \left[ 11 f_V^2 m_V^3 +
 112 B_0^2 F^2 a \right] ,
\ea
and sum of the short and long distance contributions read
\ba
\delta _2 \!\!\! & =& \!\!\!- \frac{\alpha}{\pi} \Lambda + 
\frac{3 \alpha \alpha _s}{4 \Lambda ^2} f_V^2 m_V^3 \NN \\
\delta _3\!\!\! &=& \!\!\! \frac{ \alpha \alpha _s}{12 \Lambda ^2} 
\left[ 11 f_V^2 m_V^3 + 112 B_0^2 F^2 a \right]
\rlabel{fulldelta}
\ea
We cannot simply estimate the corrections due to the quark masses here as was
done for the pseudoscalars in \rcite{dashen}. The reason is that we would
need information about the terms with 2 powers of the quark masses
for the vector meson masses. These are probably rather small since the
linear fit from section \tref{lagrangians} works rather well. This means that
we cannot estimate the quark breaking effects in the terms proportional
to $B_0 a$. We can however estimate them in the $f_V^2 m_V^3$ terms by
taking the measured values for these. This will be done in section \tref{numerics}

\section{Long distance : Meson loops}
\rlabel{mesonloops}

In this section we temporarily leave the framework of the $1/N_c$ expansion
and check whether there are any well defined contributions, nonanalytic
in the quark masses that might contribute. The mass differences of vector
mesons have contributions of order $m_q^{3/2}$\rcite{wise}. So the
electromagnetic part of the mass splittings for the pseudoscalars in fact
induces a $e^2 \sqrt{m_q}$ type of correction which is the leading quark mass
correction to the vector meson electromagnetic masses.
These come from the diagrams like \fref{diagrams}.d.
We find

\ba
\Delta \rho ^+ \!\!\! & = & \!\!\! -\frac{g^2}{6 \pi F_\pi ^2} \left\{
m_{K^+}^3 + m_{K^0}^3 + 2 m_{\pi ^+} ^3 +\frac{2}{3} m_\eta ^3 \right\}
\NN \\
\Delta \rho ^0 \!\!\! & = &  \!\!\! -\frac{g^2}{6 \pi F_\pi ^2} \left\{
m_{K^+}^3 + m_{K^0}^3 + 2 m_{\pi ^0} ^3 +\frac{2}{3} m_\eta ^3 \right\}
\NN \\
\Delta \rho \omega \!\!\! & = &  \!\!\! -\frac{g^2}{6 \pi F_\pi ^2}
\left\{ m_{K^+}^3 - m_{K^0}^3 \right \}
\NN \\
\Delta \rho \phi  \!\!\! & = &  \!\!\! 
- \frac{\sqrt{2} g^2}{6 \pi F_\pi ^2} \left\{
m_{K^+}^3 - m_{K^0}^3 \right \} 
\NN \\
\Delta \omega  \!\!\! & = & \!\!\! - \frac{g^2}{6 \pi F_\pi ^2} \left\{
m_{K^+}^3 + m_{K^0}^3 + \frac{2}{3} m_\eta^3 + 4 m_{\pi ^+}^3 +
2 m_{\pi ^0} ^3 \right\}
\NN \\
\Delta \omega \phi \!\!\! & = &  \!\!\! 
- \frac{\sqrt{2} g^2}{6 \pi F_\pi ^2} \left\{
m_{K^+}^3 + m_{K^0}^3 \right \} 
\NN \\
\Delta \phi  \!\!\! & = & \!\!\! - \frac{ g^2}{6 \pi F_\pi ^2} \left\{
2 m_{K^+}^3 + 2 m_{K^0}^3 + \frac{8}{3} m_\eta ^3 \right\}
\NN \\
\Delta K^{*+} \!\!\! & = & \!\!\! - \frac{ g^2}{6 \pi F_\pi ^2} \left\{
\frac{1}{6}m_\eta ^3 + \frac{1}{2} m_{\pi^0} ^3 + m_{\pi^+} ^3 +
2 m_{K^+}^3 + m_{K^0}^3 \right\}
\NN \\
\Delta K^{*0} \!\!\! & = & \!\!\! - \frac{ g^2}{6 \pi F_\pi ^2} \left\{
\frac{1}{6}m_\eta ^3 + \frac{1}{2} m_{\pi^0} ^3 + m_{\pi^+} ^3 +
2 m_{K^0}^3 + m_{K^+}^3 \right\}
\rlabel{pionstotal}
\ea
The  electromagnetic corrections are
\ba
\Delta ^{EM} \rho ^+ \!\!\! & = & \!\!\! -\frac{g^2}{4 \pi F_\pi ^2} \left\{
m_{K^+} + 2 m_{\pi ^+} \right\} \delta m_{P^+}^2 = -0.92 ~{\rm MeV }
\NN \\
\Delta ^{EM} \rho ^0 \!\!\! & = & \!\!\! -\frac{g^2}{4 \pi F_\pi ^2} 
m_{K^+} \delta m_{P^+}^2 = -0.59 ~{\rm MeV}
\NN \\
\Delta ^{EM} \rho \omega \!\!\! & = & \!\!\! -\frac{g^2}{4 \pi F_\pi ^2} 
m_{K^+} \delta m_{P^+}^2 = -0.59 ~{\rm MeV}
\NN \\
\Delta ^{EM} \rho \phi \!\!\! & = & \!\!\! - \frac{\sqrt{2} g^2}
{4 \pi F_\pi ^2}  m_{K^+} \delta m_{P^+}^2 = -0.83 ~{\rm MeV}
\NN \\
\Delta ^{EM} \omega \!\!\! & = & \!\!\! - \frac{g^2}{4 \pi F_\pi ^2} \left\{
m_{K^+} + 4 m_{\pi ^+} \right\} \delta m_{P ^+}^2 = -1.2 ~{\rm MeV}
\NN \\
\Delta ^{EM} \omega \phi \!\!\! & = & \!\!\! - \frac{\sqrt{2} g^2}
{4 \pi F_\pi ^2}  m_{K^+}  \delta  m_{P^+}^2 = -0.83 ~{\rm MeV}
\NN \\
\Delta ^{EM} \phi \!\!\! & = & \!\!\! - \frac{ g^2}{2 \pi F_\pi ^2} 
 m_{K^+}  \delta m_{P^+}^2  = -1.2 ~{\rm MeV}
\NN \\
\Delta ^{EM} K^{*+} \!\!\! & = & \!\!\! - \frac{ g^2}{4 \pi F_\pi ^2} \left\{
m_{\pi^+} + 2 m_{K^+} \right\} \delta m_{P^+}^2 = -1.3 ~{\rm MeV}
\NN \\
\Delta ^{EM} K^{*0} \!\!\! & = & \!\!\! - \frac{ g^2}{4 \pi F_\pi ^2} \left\{
m_{\pi^+} + m_{K^+} \right\} \delta m_{P^+}^2 = -0.75 ~{\rm MeV}
\rlabel{pionsem}
\ea
Here, $F_\pi = 93$ MeV and $\delta m_{p^+}^2=m_{\pi ^+}^2-m_{\pi ^0}^2$.
The numerical value $g=0.32$ can be obtained comparing \rref{heavyL}
with ${\cal{L}}(\rho \rho \pi)$, ( eq.(123)  of \rcite{anombij},
where VMD for the $\rho \pi \gamma$ coupling is assumed).
In \rcite{wise},
the values $g = 0.375$ (\rcite{NRQM}) and $g=0.5$ are suggested
coming from a quark meson model.
The numbers from \rref{pionstotal} and \rref{pionsem}
should not be taken too seriously since some of the total
corrections, \rref{pionstotal},  of order $O(m_q^{3/2})$, are huge:
for the $\rho$, for example, we find $\delta \rho = -223$ MeV.
In addition if terms of order $m_P\Delta m_V$,
which are formally higher order, in the expansion
of the integrals are kept there are large numerical cancellations. At that
order there are however also other contributions so those quoted above are the
only well defined ones.
For
the $\rho - \omega$ mixing, we obtain $\Delta \rho \omega = 1.8$ MeV.
 
\section{Numerical Results}
\rlabel{numerics}

We take the following inputs:
$m_s= 175$ MeV, $m_d = 8$~MeV, $m_u = 4$~MeV,
$\alpha _S$ = 0.3, $B_0= m_{K^+}^2/(m_s+m_u) =$
1360 MeV, $F_0=93$ MeV, and $m_V = (m_{\rho}, m_{K^*}, m_{\phi})$.
For $f_V$ we take 0.2 MeV when $m_V = m_\rho$ and 0.16 MeV when
$m_V = m_\phi$. These values follow from \rref{vectoree}. When
$m_V=m_{K^*}$ we take the intermediate value $f_V$ = 0.18 MeV.
It follows that $\delta_1 =(2.8,2.7,2.4)$ MeV.
From the $K^{*+} - \rho ^+$ mass difference and \rref{masses}, we get 
$a = 1.3 \times 10^{-4}$ 
MeV$^{-1}$. The parameter $d$ can be obtained from the $\rho -\omega$
mass difference. We have already seen that the $\rho ^0$, $\omega$ and
$\phi$ get mixed. The $\phi$ mixing can be ignored because the mass
splitting $m_{\phi} - m_{\omega, \rho}$ is much bigger than its mixing
and we only need to take $\rho \omega$ mixing into account. 
Because of the large $\rho$-width this contribution is also very small.
This leads to $d=(6.6,6.5,6.5)$ MeV. Next, the $\phi - K^{*0}$
mass difference implies $b=(-2.7,-2.7,-2.6) \times 10^{-6}$ MeV$^{-1}$, 
which is negligibly
small and will only be taken into account when multiplied by $m_s$.
We can attempt to obtain $\delta _2$ by combining $\Delta K^{+*}$,
$\Delta K^{0*}$, and $\Delta \rho \omega$,
finding $\delta _2 = (1,5,1.6,1.6)$ MeV.
At this point it is interesting to make a few checks of our predictions. 
For the $\rho -\phi$ mixing, \rref{masses} predicts 
$\Delta\rho \phi = (-0.67,-0.63,-0.54)$ MeV, in acceptable agreement with
\rref{expmixings}.
Notice that we also predict the sign, not given in \rref{expmixings}.
From \rref{relation} we get 
$\Delta \omega \phi = 5.4$ MeV, also in acceptable
agreement with \rref{expmixings}, but not with \rref{expmixingbis}
We can also make the prediction 
$m_{\rho ^+} - m_{\rho ^0}= (-3.0,-2.9,-2.8)$ MeV.
 It is very difficult to extract $\delta _3$. 

We could try to obtain $\Lambda$ following \rcite{dashen},
were the short and long distance contributions were matched.
However, here neither for $\delta_2$ nor for $\delta _3$ is
there a matching between the
long and short distance contributions. For $\delta _3$, there are no long
distance contributions, while for $\delta _2$ there is a disagreement even in
sign. Long and short distance
contributions do match for $\Delta \rho ^+$ and $\Delta K^{*+}$.
We find the the rather small values $\Lambda = (233,218,205)$ MeV,
which imply
$\delta _2 = (0,0.3,0.6)$ MeV and $\delta _3 = (1.5,1.9,2.3)$ MeV.

An acceptable value for $\Lambda$ should lie somewhere between
500 and 800 MeV, with
$-1.8 < \delta _2 < -1.0$ MeV for all three values of $m_V$,
and $ (0.13,0.14,0.15) < \delta _3 < (0.32,0.36,0.39)$ MeV.
Using these values we obtain for the mass differences,
$\delta V = m_{V^+} - m_{V^0}$ (MeV):
\be
(-0.4,-0.3,-0.2) < \delta ^{EM} m_{\rho} < (0.4,0.5,0.6)~
\quad\mbox{and}\quad 1.1 < \delta ^{EM} K^{*} < 1.8 .
\rlabel{deltaK}
\ee
Here $\delta ^{EM} K^{*}$ is for all three values of $m_V$.
It is very difficult to compare these numbers with the experiment,
since for the $\rho$ the present uncertainties are rather large, and the
$K^*$ receives important contributions from $m_u - m_d$.

Next, we extract the quark mass ratio
\be
-0.026 < \frac{m_u - m_d}{m_s - m_d} =
\frac{ \Delta \omega \rho - \delta _1 /6- \delta _3 /3}
{m_{K^{*+}} - m_{\rho ^+} } < -0.024,
\rlabel{quarkfrac}
\ee
in excellent agreement with \rcite{leutquarks}, who finds $-0.025$.
The $\delta_3$ contribution is small.
Using \rref{deltaK} we also obtain
\be
-0.046 < \frac{m_u - m_d}{m_s - m_d} =
\frac{ m_{K^{*+}}-m_K^{*0} +\delta_2-\delta_3/3}{m_{K^{*+}}-m_{\rho^+}}
 < -0.052 \,.
\rlabel{quarkfrac2}
\ee
These two independent estimates are in  disagreement. If we use the
value we obtained for $\delta_3$ but the value for $\delta_2$ obtained from
the phenomenological estimate $\delta_2 = 1.6 ~MeV$ the second
estimate becomes about $-0.024$ in good agreement with the other one.

The two main effects of $SU(3)$ breaking in the mass differences are 
the meson loop contributions analyzed in Section \tref{mesonloops}
and those in the photon-vector meson mixings. Both these effects we can
estimate easily. The meson loop results are listed in 
\rref{pionstotal}.
The other effect on the mass shifts we have already estimated above.
For the mixing the situation is more complicated.
Here we only need to consider $\Delta \rho \phi$ and $\Delta \omega \phi$.
This effect is not important for $\Delta \omega \phi$,
dominated by $\sqrt{2} d$ and with $\delta _1$ suppressed.
However, a momentum independent $\Delta \rho \phi$ does not make much sense.

The effects from the meson loops are rather large but not quite sufficient
to bring the estimate of \rref{quarkfrac2} in line with the standard
estimates. The expected shift for the $\rho$ and $\omega$ are
smaller.

\section{Conclusions}

We have studied corrections to vector meson masses and mixings. When the
strange quark mass is present, it dominates these corrections, the
electromagnetic ones are quite small. We  neglected corrections of order 
$O(m_s^2)$.  Our numerical predictions agree qualitatively with
the observations.

We find that even at lowest order there is no equivalent to Dashen's
theorem for vector mesons, not even in the chiral limit due to the presence
of the $\delta_3$ term which just follows from group theory.

For $m_q=0$ 
we find $\delta ^{EM} \rho ^+ = \delta ^{EM} K^{*+}$, as expected, but
$\delta ^{EM} \rho ^0 \neq \delta ^{EM} K^{*0} $, which has two sources:
vector-photon mixing, parametrized by $\delta _1$,
and short distance photons, parametrized by $\delta _3$.

We also made the prediction $ -0.4 \  < 
\delta ^{EM} m_{\rho } < 0.4$ MeV.
Here both, the $\rho ^+$ and the $\rho ^0$
receive positive mass corrections, 
the $\rho ^+$ from (long distance) photons,
and the $\rho ^0$ from the mixing with the photon.
Taking the estimate of the \rref{pionsem} as an upper limit on
the quark mass corrections, this leads to the final
prediction for the $\rho^+ - \rho^0$ mass difference:
\be
-0.7~\mbox{MeV} < m_{\rho^+}-m_{\rho^0} < 0.4~\mbox{MeV}\,.
\ee
For the $K^*$ we found using a similar estimate
\be
 0.5~\mbox{MeV} < \delta ^{EM} K^* < 1.8~\mbox{MeV}.
\ee

We finally made a prediction for a quark mass ratio in \rref{quarkfrac}
and \rref{quarkfrac2},
finding reasonable agreement with existing results, \rcite{leutquarks}
but there are indications that there are significant uncertainties in the
estimates of electromagnetic corrections to the vector meson masses.

\section{Acknowledgments}

P.G acknowledges gratefully a grant form the Spanish Ministry for
Education and Culture (Ministerio de Educaci\'on y Cultura), and
interesting and fruitful discussions with P. Talavera.

\appendix
\section{Cut-off ``heavy quark'' integrals}
The problem is that naively in dimensional regularization integrals
like
\be
I = \int \frac{d^d p}{(2\pi)^d}\quad \frac{1}{v.p + i\epsilon}
\ee
vanish. Evaluating this type of integrals with a cut-off is somewhat more 
tricky. We first rotate to euclidean space and obtain
\be
I = i\int^\Lambda\frac{ d^4 p_E }{(2\pi)^4}\frac{1}{i v.p_E + i\epsilon}\,.
\ee
here we now have to use $ 1/(x+i\epsilon) = P(1/x)-i\pi\delta(x)$. The
principal part of the integral vanishes since it is odd in $k^0_E$ and the
integration regime is symmetric. Here we specialized to $v=(1,0,0,0)$.
The remainder spatial three dimensional integral can be easily performed
and we obtain
\be
I = -\frac{i}{12\pi^2}  \Lambda^3\,.
\ee

We obtain the same result by taking the integral
\ba
I_2& =& \!\!\! \int^\Lambda \frac{d^4 q}{(2\pi)^4} \frac{1}{(p+q)^2 - m^2}\nonumber\\
 &=& \!\!\! -\frac{i}{16 \pi^2} \int_0^{\Lambda} \frac{qdq}{p_E^2}
\left( q^2+p_E^2+m^2- \sqrt{(q^2+p_E^2+m^2)^2 -4p_E^2 q^2} \right ) \\
&=& \!\!\! \frac{-i}{32 \pi ^2 p_E^2} \left\{
\frac{1}{2} \Lambda^4 + (p_E^2+m^2)\Lambda^2 -
\frac{1}{2}(\Lambda^2+m^2-p_E^2) \sqrt{ (\Lambda^2+m^2-p_E^2)^2
+ 4 m^2 p_E^2 } \right. \\
&-& \!\!\! 2 m^2 p_E^2 \log
\frac{ \Lambda^2+m^2-p_E^2+\sqrt{(\Lambda^2+m^2-p_E^2)^2+4 m^2 p_E^2}}
{m^2-p_E^2+| m^2+p_E^2| }+ \left.
\frac{1}{2} (m^2-p_E^2) |m^2+p_E^2| \right\} \nonumber
\ea

going on shell ($p_E^2 = -m^2$), and taking the limit $\Lambda$ small.
Here $p_E$ is the momentum in euclidean space.

\section{Figures}

\begin{figure}[h]
\begin{tabular}{rc}
\mbox{${ \delta M_V^{LD}}$} (MeV) &
\mbox{ $\vcenter{
\psfig{file=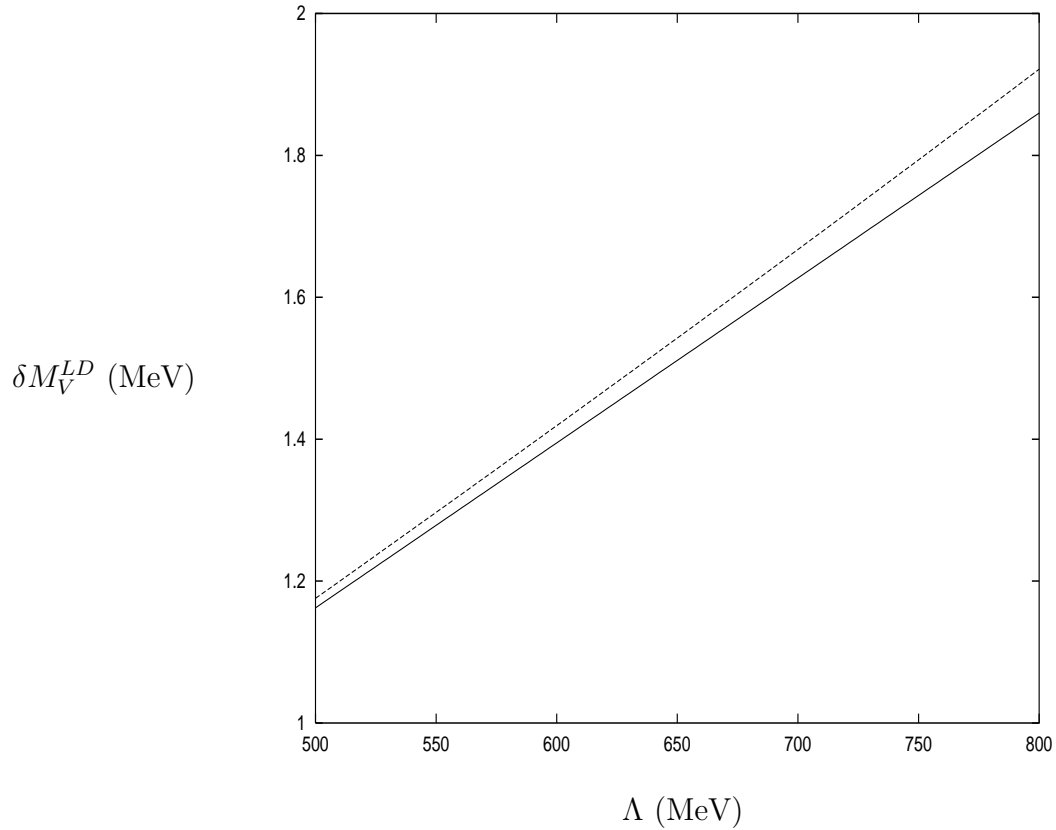,height=10cm,width=11cm,angle=270}
}$ 
} \\ &  \\ & \qquad  $\Lambda$ (MeV)
\end{tabular}
\caption{
Long distance photon correction to the charged
meson masses as a function of the cut-off. The dashed line represents the
full relativistic correction, while the straight line
represents the heavy meson effective theory prediction.
}
\rlabel{plot}
\end{figure}

\begin{figure}[h]
\psfig{figure=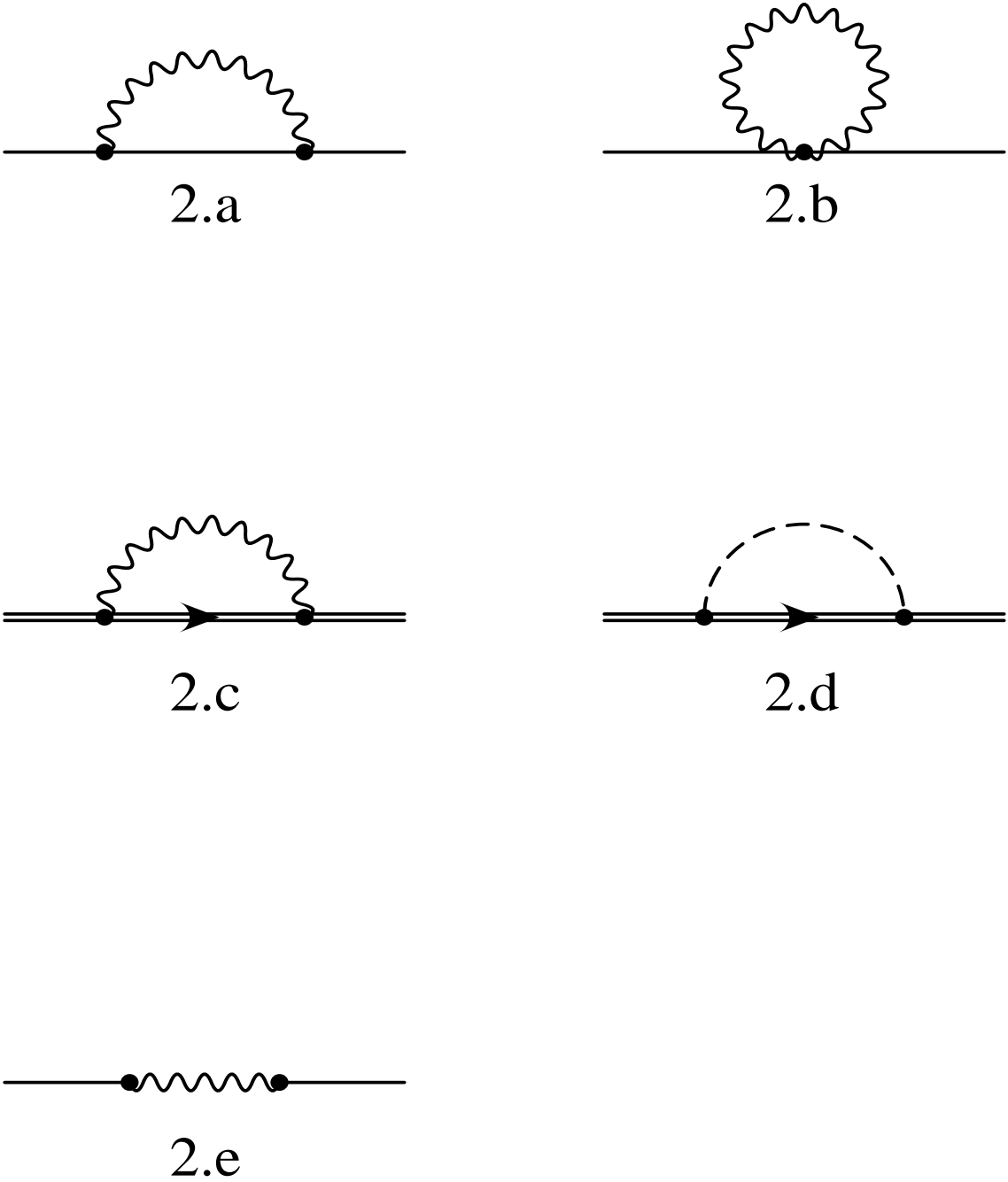,height=18cm,width=15cm}
\caption{ Feyman diagrams contributing to vector meson mass shifts and mixings.
The straight lines in 2.a and 2.b denote the mesons and the curly line
the photon. 2.b does not contribute when we using dimensional
regularization. The double lines in 2.c and 2.d denote heavy mesons.
They carry an arrow because their field destroy particles, but
do not create antiparticles. Here a diagram like 2.b 
is suppressed by $1/m_V$. The dashed line in 2.d represents a
pseudoscalar meson. The meson-photon mixing contribution is represented in 2.e.
}
\rlabel{diagrams}
\end{figure}

\end{document}